\documentstyle[twoside,fleqn]{article}
\makeatletter
\def\fileversion{v2.6}
\def\filedate{24 November 1993}

\newdimen\@bls                    
\@bls=\baselineskip               
\advance\@bls -1ex                
\newdimen\@eps                    %
\def\section{\@startsection{section}{1}{\z@}
  {1.5\@bls plus 0.5\@bls}{1\@bls}{\normalsize\bf}}
\def\subsection{\@startsection{subsection}{2}{\z@}
  {1\@bls plus 0.25\@bls}{\@eps}{\normalsize\bf}}
\def\subsubsection{\@startsection{subsubsection}{3}{\z@}
  {1\@bls plus 0.25\@bls}{\@eps}{\normalsize\bf}}
\def\paragraph{\@startsection{paragraph}{4}{\parindent}
  {1\@bls plus 0.25\@bls}{0.5em}{\normalsize\bf}}
\def\subparagraph{\@startsection{subparagraph}{4}{\parindent}
  {1\@bls plus 0.25\@bls}{0.5em}{\normalsize\bf}}

\def\@sect#1#2#3#4#5#6[#7]#8{\ifnum #2>\c@secnumdepth
  \def\@svsec{}\else
  \refstepcounter{#1}\edef\@svsec{\csname the#1\endcsname.\hskip0.5em}\fi
  \@tempskipa #5\relax
  \ifdim \@tempskipa>\z@
    \begingroup
      #6\relax
      \@hangfrom{\hskip #3\relax\@svsec}{\interlinepenalty \@M #8\par}%
    \endgroup
    \csname #1mark\endcsname{#7}\addcontentsline
      {toc}{#1}{\ifnum #2>\c@secnumdepth \else
        \protect\numberline{\csname the#1\endcsname}\fi #7}%
  \else
    \def\@svsechd{#6\hskip #3\@svsec #8\csname #1mark\endcsname
      {#7}\addcontentsline{toc}{#1}{\ifnum #2>\c@secnumdepth \else
        \protect\numberline{\csname the#1\endcsname}\fi #7}}%
  \fi \@xsect{#5}}

\long\def\@makefigurecaption#1#2{\vskip 10mm #1. #2\par}

\long\def\@maketablecaption#1#2{\hbox to \hsize{\parbox[t]{\hsize}
  {#1 \\ #2}}\vskip 0.3ex}

\def\fnum@figure{Figure \thefigure}
\def\figure{\let\@makecaption\@makefigurecaption \@float{figure}}
\@namedef{figure*}{\let\@makecaption\@makefigurecaption \@dblfloat{figure}}

\def\table{\let\@makecaption\@maketablecaption \@float{table}}
\@namedef{table*}{\let\@makecaption\@maketablecaption \@dblfloat{table}}

\textfloatsep=\floatsep           
\intextsep=\floatsep              

\long\def\@makefntext#1{\parindent 1em\noindent\hbox{${}^{\@thefnmark}$}#1}

\def\maketitle{\begingroup        
    \def\thefootnote{\fnsymbol{footnote}}%
    \newpage \global\@topnum\z@
    \@maketitle \@thanks
  \endgroup
  \let\maketitle\relax \let\@maketitle\relax
  \gdef\@thanks{}\let\thanks\relax
  \gdef\@address{}\gdef\@author{}\gdef\@title{}\let\address\relax}

\def\justify@on{\let\\=\@normalcr
  \leftskip\z@ \@rightskip\z@ \rightskip\@rightskip}

\newbox\fm@box                    

\def\@maketitle{
  \global\setbox\fm@box=\vbox\bgroup
    \vskip 8mm                    
    \raggedright                  
    \hyphenpenalty\@M             
    {\Large \@title \par}         
    \vskip\@bls                   
    {\normalsize                  
     \@author \par}               
    \vskip\@bls                   
    \@address                     
  \egroup
  \twocolumn[
    \unvbox\fm@box                
    \vskip\@bls                   
    \unvbox\abstract@box          
    \vskip 2pc]}                  

\newcounter{address}
\def\theaddress{\alph{address}}
\def\@makeadmark#1{\hbox{$^{\rm #1}$}}

\def\address#1{\addressmark\begingroup
  \xdef\@tempa{\theaddress}\let\\=\relax
  \def\protect{\noexpand\protect\noexpand}\xdef\@address{\@address
  \protect\addresstext{\@tempa}{#1}}\endgroup}
\def\@address{}

\def\addressmark{\stepcounter{address}%
  \xdef\@tempb{\theaddress}\@makeadmark{\@tempb}}

\def\addresstext#1#2{\leavevmode \begingroup
  \raggedright \hyphenpenalty\@M \@makeadmark{#1}#2\par \endgroup
  \vskip\@bls}

\newbox\abstract@box              

\def\abstract{%
  \global\setbox\abstract@box=\vbox\bgroup
  \small\rm
  \ignorespaces}
\def\endabstract{\par \egroup}

\def\thebibliography#1{\section*{REFERENCES}\list{\arabic{enumi}.}
  {\settowidth\labelwidth{#1.}\leftmargin=1.67em
   \labelsep\leftmargin \advance\labelsep-\labelwidth
   \itemsep\z@ \parsep\z@
   \usecounter{enumi}}\def\makelabel##1{\rlap{##1}\hss}%
   \def\newblock{\hskip 0.11em plus 0.33em minus -0.07em}
   \sloppy \clubpenalty=4000 \widowpenalty=4000 \sfcode`\.=1000\relax}

\newcount\@tempcntc
\def\@citex[#1]#2{\if@filesw\immediate\write\@auxout{\string\citation{#2}}\fi
  \@tempcnta\z@\@tempcntb\m@ne\def\@citea{}\@cite{\@for\@citeb:=#2\do
    {\@ifundefined
       {b@\@citeb}{\@citeo\@tempcntb\m@ne\@citea
        \def\@citea{,\penalty\@m\ }{\bf ?}\@warning
       {Citation `\@citeb' on page \thepage \space undefined}}%
    {\setbox\z@\hbox{\global\@tempcntc0\csname b@\@citeb\endcsname\relax}%
     \ifnum\@tempcntc=\z@ \@citeo\@tempcntb\m@ne
       \@citea\def\@citea{,\penalty\@m}
       \hbox{\csname b@\@citeb\endcsname}%
     \else
      \advance\@tempcntb\@ne
      \ifnum\@tempcntb=\@tempcntc
      \else\advance\@tempcntb\m@ne\@citeo
      \@tempcnta\@tempcntc\@tempcntb\@tempcntc\fi\fi}}\@citeo}{#1}}

\def\@citeo{\ifnum\@tempcnta>\@tempcntb\else\@citea
  \def\@citea{,\penalty\@m}%
  \ifnum\@tempcnta=\@tempcntb\the\@tempcnta\else
   {\advance\@tempcnta\@ne\ifnum\@tempcnta=\@tempcntb \else
\def\@citea{--}\fi
    \advance\@tempcnta\m@ne\the\@tempcnta\@citea\the\@tempcntb}\fi\fi}

\def\ps@crcplain{\let\@mkboth\@gobbletwo
     \def\@oddhead{\reset@font{\sl\rightmark}\hfil \rm\thepage}%
     \def\@evenhead{\reset@font\rm \thepage\hfil\sl\leftmark}%
     \let\@oddfoot\@empty
     \let\@evenfoot\@oddfoot}

\ps@crcplain                    

\newcommand{\AmS}{{\protect\the\textfont2
  A\kern-.1667em\lower.5ex\hbox{M}\kern-.125emS}}

\makeatother

\newcommand{\bee}{\begin{equation}}
\newcommand{\ene}{\end{equation}}

\title{A Newtonian Model for the Quantum Gravitational Back-Reaction
on Inflation}

\author{{Richard P. Woodard}
\address{Department of Physics, University of Florida,
Gainesville, FL 32611, United States}%
\thanks{Supported by DOE contact DE-FG02-97ER-41029 and by the
Institute for Fundamental Theory.}}

\begin{document}
\begin{abstract}
Quantum gravitational back-reaction offers a simultaneous explanation
for why the cosmological constant is so small and a natural model of
inflation in which scalars play no role. In this talk I review
previous work and present a simple model of the mechanism in which 
the induced stress tensor behaves like negative vacuum energy with a 
density proportional to $-\Lambda/{8\pi G} \cdot (G \Lambda)^2 \cdot H t$. 
The model also highlights the essential role of causality in back-reaction.
\end{abstract}

\maketitle

The application of a force field in quantum field theory generally
rearranges virtual quanta and thereby induces currents and/or stresses
which modify the original force field. This is the phenomenon of {\it
back-reaction}. Famous examples include the response of QED to a
homogeneous electric field [1] and the response of generic
matter theories to the gravitational field of a black hole [2].
In the former case, virtual $e^+ e^-$ pairs can acquire the energy
needed to become real by tunnelling up and down the field lines. The
newly created pairs are also accelerated in the electric field, which
gives a current that reduces the original electric field. The event
horizon of a black hole also causes particle creation when one member
of a virtual pair passes out of causal contact with the other by
entering the event horizon. As the resultant Hawking radiation carries
away the black hole's mass the surface gravity rises.

Parker [3] was the first to realize that the expansion of
spacetime can lead to the production of massless, minimally coupled
scalars. Grishchuk [4] later showed that the same mechanism
applies to gravitons. Production of these particles is especially
efficient during inflation because of the causal horizon. Consider a 
spatially flat, locally de Sitter geometry with Hubble constant $H$, 
the invariant element for which is,
\begin{equation}
ds^2 = -dt^2 + e^{2 H t} d{\vec x} \cdot d{\vec x} \; . \label{(1)}
\end{equation}
Whereas the co-moving time $t$ corresponds to the duration of
physical processes, the physical distance between two points
${\vec x}$ and ${\vec y}$ is not their coordinate separation,
${\Delta \ell} \equiv \Vert {\vec x} - {\vec y}\Vert$, but rather
$e^{H t} {\Delta \ell}$. This is a geometry in which Zeno's paradox
about the impossibility of motion can come true! For if a photon 
is emitted at time $t=0$ from an observer more distant than one 
Hubble length, $H^{-1}$, then the observer will never see it. By
the time the photon has covered half the initial distance, the
spacetime in between will have expanded by so much that the photon
actually has {\it further} to go.

Now add the Uncertainty Principle. Even in flat space this requires
virtual particles to be continually emerging from the vacuum and then
disappearing back into it. But massless particles which are not also 
conformally invariant have a reasonable amplitude for appearing with 
wavelength greater than $H^{-1}$. When this happens in an inflating 
universe the particles must be ripped apart from one another by the 
Hubble flow. Since this creation mechanism requires both effective 
masslessness on the scale of inflation and the absence of conformal 
invariance, it is limited to gravitons and to light, minimally 
coupled scalars. 

There is no doubt as to the reality of inflationary particle
production because it is the usual explanation [5,6] for the
primordial spectrum of cosmological density perturbations whose 
imprint on the cosmic microwave background has been so clearly
imaged by the latest balloon experiments [7,8]. Our special interest 
is the back-reaction from this process. There is no buildup of 
particle density because the 3-volume expands as new particles are
created so as to keep the density constant. When a new pair is pulled
out of the vacuum the one before it is, on average, already in another
Hubble volume. However, the gravitational field is another thing. 
The created particles are highly infrared so they do not carry much 
stress-energy, but they do carry some, and this must engender a
gravitational field in the region between them. Because gravity is
attractive this field acts to resist the Hubble flow. Further, it is
cumulative. The gravitational field of a created pair must remain 
behind to add with that of subsequent pairs, just as an observer 
continues to feel the gravity of a pebble even after dropping it into 
a black hole. 

We believe that the gravitational attraction between virtual infrared 
gravitons gradually builds up a restoring force that impedes further 
inflation [9]. Gravity is a weak interaction, even at the enormous
energies usually conceived for inflation, so the increment from each 
new pair is minuscule. But the effect must continue to grow until it 
becomes non-perturbatively strong, unless something else supervene to 
end inflation first. This mechanism offers the dazzling prospect of 
simultaneously resolving the (old) problem of the cosmological constant 
[10] and providing a natural model of inflation in which there is no 
scalar inflaton. The idea is that the actual cosmological constant is 
not small and that this is what caused inflation during the early 
universe. Back-reaction plays the crucial role of ending inflation.

A sometimes confusing point is that one does not require the 
complete theory of quantum gravity in order to study an infrared 
process such as this. As long as spurious time dependence is not 
injected through the ultraviolet regularization, the late time 
back-reaction is dominated by ultraviolet finite, nonlocal terms 
whose form is entirely controlled by the low energy limiting theory. 
This theory must be general relativity, 
\begin{equation}
{\cal L} = {1 \over 16 \pi G} (R - 2 \Lambda) \sqrt{-g} \; ,
\label{(2)}
\end{equation}
with the possible addition of some light scalars. Here ``light'' 
means massless with respect to $H \equiv \sqrt{\Lambda/3}$. No other 
quanta can contribute effectively in this regime.

It is worth commenting that infrared phenomena can always be studied 
using the low energy effective theory. This is why Bloch and Nordsieck 
[11] were able to resolve the infrared problem of QED before the 
theory's renormalizability was suspected. It is also why Weinberg [12] 
was able to achieve a similar resolution for $\Lambda = 0$ quantum 
general relativity. And it is why Feinberg and Sucher [13] were able 
to compute the long range force due to neutrino exchange using Fermi 
theory. More recently Donoghue [14] has been working along the same 
lines for $\Lambda = 0$ quantum gravity.

To see a crude, Newtonian version of the process, consider locally de 
Sitter inflation on the manifold $T^3 \times R$, so that the physical 
radius of the universe at co-moving time $t$ is,
\begin{equation}
r(t) \sim H^{-1} e^{H t} \; . \label{(3)}
\end{equation}
The {\it bare} energy density of inflationally produced infrared 
gravitons --- just the $\hbar \omega$ per graviton --- is [4],
\begin{equation}
\rho_{\rm IR} \sim H^4 \; . \label{(4)}
\end{equation}
This is insignificant compared with the energy density of the
cosmological constant, $\Lambda/{8\pi G}$, and $\rho_{\rm IR}$ is in 
any case positive. However, the gravitational interaction energy is 
negative, and it can be enormous if there is contact between a large 
enough fraction of the total mass of infrared gravitons,
\begin{equation}
M(t) \sim r^3(t) \rho_{\rm IR} \sim H e^{3 H t} \; . \label{(5)}
\end{equation}
For example, if $M(t)$ was {\it all} in contact with itself the Newtonian
interaction energy would be,
\begin{equation}
- {G M^2(t) \over r(t)} \sim -G H^3 e^{5 H t} \; , \label{(6)}
\end{equation}
Dividing by the 3-volume gives a density of about $-G H^6 e^{2 H t}$, which
rapidly becomes enormous.

Of course this ignores causality. Most of the infrared gravitons needed to
maintain $\rho_{\rm IR}$ are produced out of causal contact with one another
in different Hubble volumes. The ones in gravitational interaction are those
produced within the same Hubble volume. Since the number of Hubble volumes
grows like $e^{3 H t}$, the rate at which mass is produced within a single
Hubble volume is,
\begin{equation}
{d M_1 \over dt} \sim H^2 \; . \label{(7)}
\end{equation}
Although most of the newly produced gravitons soon leave the Hubble volume,
their gravitational potentials must remain, just as an outside observer
continues to feel the gravity of particles that fall into a black hole. The
rate at which the Newtonian potential accumulates is therefore,
\begin{equation}
{d \Phi_1 \over dt} \sim - {G \over H^{-1}} {d M_1 \over dt} 
\sim -G H^3 \; . \label{(8)}
\end{equation}
Hence the Newtonian gravitational interaction energy density is,
\begin{eqnarray}
\rho(t) & \sim & \rho_{\rm IR} \Phi_1(t) \sim -G H^6 H t \label{(9)} \\
& \sim & - {\Lambda \over 8\pi G} \cdot (G \Lambda)^2 \cdot H t 
\; . \label{(10)}
\end{eqnarray}

When the number of e-foldings $Ht$ becomes large, the fractional rate of change
of the gravitational interaction energy is negligible compared with the
expansion rate,
\begin{equation}
\vert {\dot \rho}(t) \vert \ll H \vert \rho(t) \vert \; . \label{(11)}
\end{equation}
It follows from energy conservation,
\begin{equation}
{\dot \rho}(t) = -3 H \Bigl(\rho(t) + p(t)\Bigr) \; , \label{(12)}
\end{equation}
that the induced pressure must be nearly opposite to the energy density. In
other words, back-reaction induces negative vacuum energy.

If inflation persists long enough, the minuscule dimensionless coupling
constant in (10) --- $(G \Lambda)^2$ --- can be overcome by the secular
growth in the number of e-foldings, $H t$. Of course nonlinear effects become
important as well. The physical picture is of inflation ending, rather
suddenly, with the universe poised on the verge of gravitational collapse.

Detailed, explicit computations confirm the qualitative analysis 
presented above, at two loops for pure gravity [15] and at three loops for
a massless, minimally coupled $\phi^4$ theory in which the scalar
self-interaction provides the breaking mechanism [16]. We worked on
the manifold $T^3 \times R$ and used the Schwinger-Keldysh formalism
[17] to compute the expectation value of the metric in the presence
of a state which is free, Bunch-Davies vacuum at $t=0$. Since the
state is homogeneous and isotropic the expectation value of the
metric can be put in the form,
\begin{equation}
\langle \Omega \vert g_{\mu\nu}(t,\vec{x}) dx^{\mu} dx^{\nu} \vert 
\Omega \rangle = -dt^2 + e^{2 b(t)} d{\vec x} \cdot d{\vec x} 
\; . \label{(13)}
\end{equation}
We used perturbation theory about the background $b_0(t) = H t$.
There is no secular slowing at one loop because the inflationary
production of gravitons is itself a one loop effect and the
back-reaction from it must come at a higher order in perturbation
theory. Including two loop effects gives the following result for
the expansion rate,
\begin{eqnarray}
\lefteqn{\dot{b}(t) = H \left\{ 1 - \left({G \Lambda \over 3\pi}\right)^2 
\left[\frac16 (H t)^2 \right. \right. } \nonumber \\
& & \left. \left. {\mbox{} \over \mbox{}}  + O(Ht) \right] + 
O(G^3)\right\} \; . \label{(14)}
\end{eqnarray}
The extra factor of $Ht$ not present in our Newtonian model derives
from general relativistic effects having to do with the response to
pressure in a locally de Sitter background. 

Although we did not compute contributions at three loops and higher we 
were able to obtain all orders bounds on their time dependence. These 
bounds show that the two loop result dominates for as long as 
perturbation theory remains valid [18]. Of course one cannot trust this 
analysis when the effect becomes of order one. The correct statement 
is that back-reaction slows inflation by an amount that eventually 
becomes nonperturbatively large. We have some ideas [19] about how to 
evolve beyond this point but no firm results as yet.

It is highly significant that this model of inflation contains only
a single free parameter, the dimensionless product of Newton's
constant and the cosmological constant, $G \Lambda$. This means that 
the model makes unique and testable predictions in a way that
generic scalar-driven inflation can never do. In fact the model's
predictions for the CMBR parameters $A_S$, $r$, $n_S$ and $n_T$ 
have recently been worked out [20]. The scalar amplitude $A_S$ sets
$G \Lambda$, but the other three are legitimate predictions of the
model. They are in agreement with the latest data and will be subject
to much closer experimental scrutiny from the precision satellite
probes which are due to be flown in the coming decade. These may be
the first observations of quantum gravitational phenomena!


\begin{thebibliography}{9}
\bibitem{1} 
J. Schwinger, Phys. Rev. {\bf 89} (1951) 664.
\bibitem{2}
S. W. Hawking, Nature {\bf 248} (1974) 30;
Commun. Math. Phys. {\bf 43} (1975) 199.
\bibitem{3}
L. Parker, Phys. Rev. Lett. {\bf 21} (1968) 562;
Phys. Rev. {\bf 183} (1969) 1057.
\bibitem{4} 
L. P. Grishchuk, Sov. Phys. JETP {\bf 40} (1974) 409.
\bibitem{5}
V. F. Mukhanov, H. A. Feldman and R. H. Brandenberger, 
Phys. Rep. {\bf 215} (1992) 203.
\bibitem{6}
A. R. Liddle and D. H. Lyth, Phys. Rep. {\bf 231} (1993) 1.
\bibitem{7}
P. de Bernardis et al., Nature {\bf 404} (2000) 995.
\bibitem{8}
S. Hanany et al., Astrophys. J. Lett. {\bf 545} (2000) 5.
\bibitem{9}
N. C. Tsamis and R. P. Woodard, Nucl. Phys {\bf B474} (1996) 235.
\bibitem{10} 
S. M. Carroll, Living Rev. Rel. {\bf 4} (2001) 1.
\bibitem{11} 
F. Bloch and H. Nordsieck, Phys. Rev. {\bf 52} (1937) 54.
\bibitem{12} 
S. Weinberg, Phys. Rev. {\bf 140} (1965) B516.
\bibitem{13} 
G. Feinberg and J. Sucher, Phys. Rev. {\bf 166} (1968) 1638.
\bibitem{14} 
J. F. Donoghue, Phys. Rev. {\bf D50} (1994) 3874;
Phys. Rev. Lett. {\bf 72} (1994) 2996.
\bibitem{15}
N. C. Tsamis and R. P. Woodard, Ann. Phys. {\bf 253} (1997) 1.
\bibitem{16}
N. C. Tsamis and R. P. Woodard, Phys. Lett. {\bf B426} (1998) 21.
\bibitem{17} 
J. Schwinger, J. Math. Phys. {\bf 2} (1961) 407.
\bibitem{18}
N. C. Tsamis and R. P. Woodard, Ann. Phys. {\bf 267} (1998) 145.
\bibitem{19}
N. C. Tsamis and R. P. Woodard, Phys. Rev. {\bf D57} (1998) 4826.
\bibitem{20}
L. R. Abramo, N. C. Tsamis and R. P. Woodard, Fortschritte der
Physik {\bf 47} (1999) 389.
\end{thebibliography}
\end{document}